\documentclass[12pt]{JHEP} 
\usepackage{graphicx}
\newcommand\fverb{\setbox\pippobox=\hbox\bgroup\verb}
\newcommand\fverbdo{\egroup\medskip\noindent%
			\fbox{\unhbox\pippobox}\ }
\newcommand\fverbit{\egroup\item[\fbox{\unhbox\pippobox}]}
\newbox\pippobox

\def\del{\partial}

\def\H{{\cal H}}

\def\psid{\psi^\dagger}

\def\ad{a^\dagger}

\def\ads{AdS$_5 \times$S$^5$}

\def\be{\begin{equation}}
\def\ee{\end{equation}}
\def\ba{\begin{array}{l}}
\def\ea{\end{array}}
\def\bea{\begin{eqnarray}}
\def\eea{\end{eqnarray}}
\def\beas{\begin{eqnarray*}}
\def\eeas{\end{eqnarray*}}
\def\eq#1{(\ref{#1})}

\def\nn{\nonumber\\}

\def\ket#1{| #1 \rangle}
\def\bra#1{ \langle #1 |}

\def\psid{\psi^\dagger}

\def\ad{a^\dagger}
\def\sigmad{\sigma^\dagger}

\def\ads{$AdS_5 \times S^5$}

\title{Bosonization of non-relativistic fermions on a circle: Tomonaga's problem revisited}
\author{Avinash Dhar and Gautam Mandal \\  
Tata Institute of Fundamental Research, Homi Bhabha Road, \\
Mumbai 400 005, India 
\\~~\\
\email{adhar@theory.tifr.res.in, mandal@theory.tifr.res.in
}}

\preprint{\hepth{0603154}\\
TIFR/TH/06-08
}
     
\abstract
{We use the recently developed tools for an exact bosonization of a
finite number $N$ of non-relativistic fermions to discuss the classic
Tomonaga problem. In the case of noninteracting fermions, the
bosonized hamiltonian naturally splits into an O$(N)$ piece and an
O$(1)$ piece. We show that in the large-$N$ and low-energy limit, the
O$(N)$ piece in the hamiltonian describes a massless relativistic
boson, while the O$(1)$ piece gives rise to cubic self-interactions of
the boson. At finite $N$ and high energies, the low-energy effective
description breaks down and the exact bosonized hamiltonian must be
used. We also comment on the connection between the Tomonaga problem and
pure Yang-Mills theory on a cylinder. In the dual context  
of baby universes and multiple black holes in string theory, we point out 
that the O$(N)$ piece in our bosonized hamiltonian provides a simple 
understanding of the origin of two different kinds of nonperturbative
O$(e^{-N})$ corrections to the black hole partition function.}


\begin{document}  
    

\section{Introduction}

In a recent work \cite{DMS} we have discussed an exact bosonization of
a finite number $N$ of non-relativistic fermions. The tools developed
there have many potential applications to several areas of
physics. An example is the half-BPS sector of ${\cal N}=4$ super
Yang-Mills theory and its holographic dual string theory in \ads~
space-time, some aspects of which were discussed in \cite{DMS} and
more recently in greater detail in \cite{DMM}. In the present work we
will discuss application of our exact bosonization to the classic
Tomonaga problem \cite{Tomonaga} of non-relativistic fermions on a
circle.

Historically, the Tomonaga problem has played an important 
role\footnote{See, for example, \cite{S} for an account of this.} in the 
development of tools for treating a system of large number of fermions
interacting with long-range forces like Coulomb force, which is an
essential problem one encounters in condensed matter systems. In the
toy model of a $1$-dimensional system of non-relativistic fermions, 
Tomonaga was the first to show that interactions can
mediate new collective dynamical degrees of freedom, which are
quantized as bosons. The essential idea was the observation that a
long-range (in real space) force like Coulomb interaction becomes
short range in momentum space, so particle-hole pairs in a low-energy
band around fermi surface (which  consists of just two points in $1$-dimension)
do not get ``scattered out'' of the band. As a result, the interacting
ground state, as well as excited states with low excitation energy
compared to the fermi energy, involve only particle-hole pairs in a
small band around the fermi surface. For a large number of fermions,
there is a finite band around the fermi surface in which the excited
states satisfy this requirement. In his work, Tomonaga laid down
precise criteria for the collective boson low-energy approximation to
work, and showed that when his criteria are met, both the free as well
as the interacting fermion systems can be described by a system of
free bosons, under a suitable approximation, for a system of large
number of fermions.

In the present work, we will apply our exact bosonization tools to the
Tomonaga problem. The states of our bosonized theory are multiparticle
states of a system of free bosons, each of which can be in any of the
first $N$ levels of a harmonic oscillator, where $N$ is the number of
fermions. These states diagonalize the non-interacting part of the
fermion hamiltonian exactly. We will show that the standard effective
low-energy theory for appropriate Fourier modes of the spatial fermion
density operator can be derived from this bosonized theory. We will
see that the conditions under which this can be done are precisely the
ones required by Tomonaga for his approximations to work. However, for
the non-interacting case \footnote{In the case of interacting
fermions, our bosons are generally interacting and then approximations
become necessary to make further progress. For example, a four-fermi
interaction may be possible to handle in the limit of a large number
of fermions at low-energies. This is discussed further in Sec 6.} 
our exact bosonization is applicable even outside the regime of
validity of the low-energy approximation.

The organization of this paper is as follows. In Sec 2 we summarize
the work of \cite{DMS} \footnote{The work in this paper discusses two
different exact bosonizations of the fermi system; here we will limit
our discussion to bosonization of the first type.} on the exact
bosonization of a finite number $N$ of nonrelativistic fermions. The
presentation here is different and simpler, though completely
equivalent to that in \cite{DMS}. The merit of this presentation is
that it makes the bosonization rules simpler and graphical, making
applications of these rules very easy. In Sec 3 we derive the
bosonized hamiltonian for a system of $N$ non-relativistic fermions on
a circle. The free fermion system has a degenerate
spectrum, so we need to do further work before applying the
bosonization rules of Sec 2. After explaining how to take care of the
degeneracy, we derive the bosonized form of the hamiltonian. The
nontrivial part of the bosonized hamiltonian naturally splits into a
sum of a large, O$(N)$, piece and a small, O$(1)$ piece. We show that
ignoring the latter corresponds to the relativistic boson
approximation of Tomonaga. We do this by computing the partition
function of the O$(N)$ piece in the hamiltonian, which reduces at
large $N$ to that of a relativistic boson. The O$(1)$ piece
gives rise to a cubic self-interaction of this boson. This is shown in Sec
4 where we derive the effective low-energy cubic hamiltonian for these
self-interactions. Extension beyond low-energy approximation is
discussed in Sec 5. Interacting fermion case is discussed in Sec
6. Connection of the Tomonaga problem with Yang-Mills theory on a
two-dimensional cylinder is discussed in Sec 7. We end with a summary
and some comments in Sec 8. Details of some computations described in
the text are given in Appendices A, B and C. In Appendix D we discuss
possible extension of our bosonization techniques to higher
dimensions.

\section{Review of exact bosonization}

In this section we will review the techniques developed in \cite{DMS}
for an exact operator bosonization of a finite number of
nonrelativistic fermions. The discussion here is somewhat different
from that in \cite{DMS}. Here, we will derive the first bosonization
of \cite{DMS} using somewhat simpler arguments, considerably
simplifying the presentation and the formulae in the
process. Moreover, the present derivation of bosonization rules is
more intuitive, making its applications technically easier.

Consider a system of $N$ fermions each of which can occupy
a state in an infinite-dimensional Hilbert space  $\H_f$.
Suppose there is a countable basis of $\H_f:
\{ \ket{m}, m=0,1, \cdots, \infty\}$. For example, this could
be the eigenbasis of a single-particle hamiltonian, $\hat h \ket{m} =
{\cal E}(m) \ket{m}$, although other choices of basis would do equally
well, as long as it is a countable basis. In the second quantized
notation we introduce creation (annihilation) operators
$\psi^\dagger_m$ ($\psi_m$) which create (destroy) particles in the
state $\ket{m}$. These satisfy the anticommutation relations
\be
\{ \psi_m, \psid_n\}= \delta_{mn}
\label{anticom}
\ee
The $N$-fermion states are  given by (linear combinations of)
\be
\ket{f_1, \cdots, f_N}= \psid_{f_1}\psid_{f_2} \cdots
\psid_{f_N} \ket{0}_F
\label{fermi-state}
\ee
where $f_m$ are arbitrary integers satisfying 
$0\le f_1 < f_2 < \cdots < f_N$, and 
$\ket{0}_F$ is the usual 
Fock vacuum annihilated by $\psi_m, m=0,1, \cdots, \infty$.

It is clear that one can span the entire space of $N$-fermion states,
starting from a given state $\ket{f_1, \cdots, f_N}$, by repeated
application of the fermion bilinear operators
\be
\Phi_{mn} = \psid_m \, \psi_n
\label{phi-mn}
\ee
However, the problem with these bosonic operators is that they are not
independent; this is reflected in the W$_\infty$ algebra that they
satisfy,
\be
[\Phi_{mn}, \Phi_{m'n'}] = \delta_{m'n}\Phi_{mn'} - \delta_{mn'}\Phi_{m'n}.
\label{winf}
\ee
This is the operator version of the noncommutative constraint $u*u=u$
that the Wigner distribution $u$ satisfies in the exact path-integral
bosonization carried out in \cite{DMW-nonrel}.

A new set of unconstrained bosonic operators was introduced in
\cite{DMS}, $N$ of them for $N$ fermions. In effect, this set of
bosonic operators provides the independent degrees of freedom in terms
of which the above constraint is solved. Let us denote these operators
by $\sigma_k,~k=1, 2, \cdots, N$ and their conjugates,
$\sigmad_k,~k=1, 2, \cdots, N$. As we shall see shortly, these
operators will turn out to be identical to the $\sigma_k$'s used in
\cite{DMS}. The action of $\sigmad_k$ on a given fermion state
$\ket{f_1, \cdots, f_N}$ is rather simple.
It just takes each of the fermions in the top $k$ occupied levels 
up by one step, as illustrated in Figure 1.
One starts
from the fermion in the topmost occupied level, $f_N$, and moves it up
by one step to $(f_N+1)$, then the one below it up by one step, etc
proceeding in this order, all the way down to the $k$th fermion from
top, which is occupying the level $f_{N-k+1}$ and is taken to the
level $(f_{N-k+1}+1)$. For the conjugate operation, $\sigma_k$, one
takes fermions in the top occupied $k$ levels down by one step,
reversing the order of the moves. Thus, one starts by moving the
fermion at the level $f_{N-k+1}$ to the next level below at
$(f_{N-k+1}-1)$, and so on. Clearly, in this case the answer is nonzero
only if the $(k+1)$th fermion from the top is not occupying the level
immediately below the $k$th fermion , i.e. only  if $(f_{N-k+1}-f_{N-k}-1) >
0$. If $k=N$ this condition must be replaced by $f_1 > 0$.
     \begin{figure}[htb]
       \centering
       \includegraphics[height=7cm]{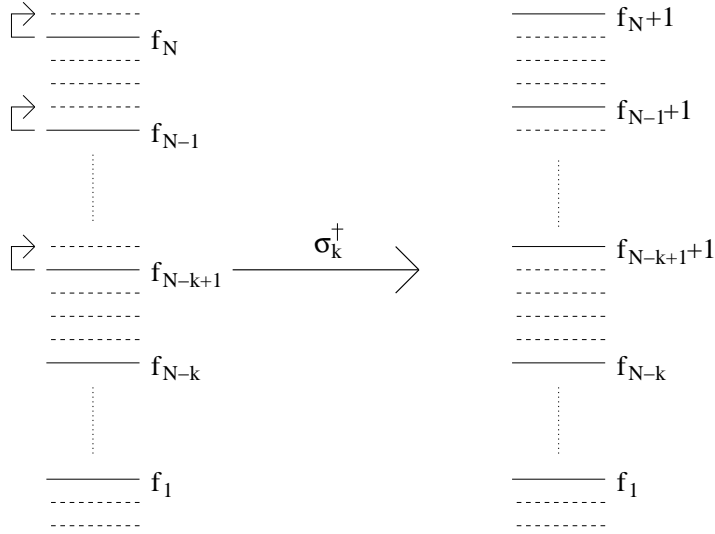}
       \caption{The action of $\sigma_{k}^{\dagger}$.}
       \label{fig:1}
     \end{figure}

These operations are necessary and sufficient to move to any desired
fermion state starting from a given state. This can be argued as
follows. First, consider the following operator,
$\sigma_{k-1}~\sigmad_k$.  Acting on an arbitrary fermion state, the
first factor takes top $k$ fermions up by one level; this is followed
by bringing the top $(k-1)$ fermions down by one level. The net effect
is that only the $k$th fermion from top is moved up by one level. In
other words,
$\sigma_{k-1}~\sigmad_k=\psid_{f_{N-k+1}+1}~\psi_{f_{N-k+1}}=
\Phi_{{f_{N-k+1}+1},f_{N-k+1}}$. In this way, by composing together
different $\sigma_k$ operations we can move individual fermions
around. Clearly, all the $N$ $\sigma_k$ operations are necessary in
order to move each of the $N$ fermions indvidually. It is easy to see
that by applying sufficient number of such fermion bilinears one can
move to any desired fermion state starting from a given state.

It follows from the definition of $\sigmad_k,~\sigma_k$ operators that 
they satisfy the following relations:
\bea
\sigma_k~\sigmad_k= 1,~~\sigmad_k~\sigma_k=\theta(r_k-1),~~
[\sigma_l, \sigmad_k]=0,~~l \neq k,
\label{sigmas}
\eea
where $(f_{N-k+1}-f_{N-k}-1) \equiv r_k$ and $\theta(m)=1$ if $m \geq
0$, otherwise it vanishes. Moreover, all the $\sigma_k$'s annihilate the
Fermi vacuum.

Consider now a system of bosons each of which can occupy a state in an
$N$-dimensional Hilbert space $\H_N$. Suppose we choose a basis
$\{\ket{k},~k=1,\cdots,N\}$ of $\H_N$. In the second quantized
notation we introduce creation (annihilation) operators $a^\dagger_k$
($a_k$) which create (destroy) particles in the state $\ket{k}$. These
satisfy the commutation relations
\be
[a_k, \ad_l]= \delta_{kl}, \quad k,l=1, \cdots, N
\label{oscillator}
\ee 
A state of this bosonic system is given by (a linear combination
of)
\be
\ket{r_1, \cdots, r_N}= \frac{(a_1^\dagger)^{r_1}\cdots
(a_N^\dagger)^{r_N}}{ \sqrt{r_1 ! \cdots r_N!}} \ket{0}
\label{bose-state}
\ee

It can be easily verified that equations (\ref{sigmas}) are 
satisfied if we make the following identifications
\bea
\sigma_k &=& \frac{1}{\sqrt{\ad_k a_k + 1}} a_k, \nn
\sigmad_k &=& \ad_k\frac{1}{\sqrt{\ad_k a_k + 1}},
\label{sigmadefs}
\eea
together with the map
\bea
r_k &=& f_{N-k+1} - f_{N-k} - 1, \quad \quad k = 1,\,2,\,\cdots\,N-1 \nn
r_N &=& f_1.
\label{statemap} 
\eea
This identification is consistent with the Fermi vacuum being the
ground state of the bosonic system. The map \eq{statemap} first
appeared in \cite{Nemani}. The first of these arises from the
identification (\ref{sigmadefs}) of $\sigma_k$'s in terms of the
oscillator modes, while the second follows from the fact that
$\sigma_N$ annihilates any state in which $f_1$ vanishes.

Using the above bosonization formulae, any fermion bilinear operator
can be expressed in terms of the bosons. For example, the hamiltonian
can be rewritten as follows. Let ${\cal E}(m),~m=0, 1, 2, \cdots$ be
the exact single-particle spectrum of the fermions (assumed
noninteracting). Then, the hamiltonian is given by 
\be 
H = \sum_{m=0}^\infty {\cal E}(m) ~ \psid_m~\psi_m.
\label{fermiham}
\ee
Its eigenvalues are $E=\sum_{k=1}^{N} {\cal E}(f_k)$. 
Using $f_k=\sum_{i=N-k+1}^{N} r_i +k-1$, which is easily derived from 
(\ref{statemap}), these can be rewritten in terms of the bosonic occupation
numbers, $E=\sum_{k=1}^{N} {\cal E}(\sum_{i=N-k+1}^{N} r_i +k-1)$. These 
are the eigenvalues of the bosonic hamiltonian
\be
H = \sum_{k=1}^{N} {\cal E}(\hat n_k), \quad 
\hat n_k \equiv \sum_{i=k}^{N} \ad_i a_i +N-k
\label{boseham}
\ee
This bosonic hamiltonian is, of course, completely equivalent to the 
fermionic hamiltonian we started with.

We end this section with the remark that our bosonization technique
does not depend on any specific choice of fermion hamiltonian and can
be applied to various problems like $c=1$, half-BPS sector of ${\cal
N}=4$ super Yang-Mills theory \cite{DMM}, etc.

\section{Free non-relativistic fermions on a circle - Large $N$ limit}

In this and the next section, we will discuss the theory obtained by
bosonization of the noninteracting part of the fermion hamiltonian. 
The case of interacting fermions will
be taken up in Sec 5. In the second quantized langauge, we may write
the free part as
\be
H_{\rm free} = -\frac{\hbar^2}{2m}\int_0^L dx \ \chi^\dagger(x) \del_x^2 \chi(x).
\label{fermi-free1}
\ee
Here $L$ is the size of the circle and $m$ is the mass of each
fermion. In terms of the fourier modes, $\chi_{\pm n} \equiv
\int_0^L \frac{dx}{\sqrt{L}} \ e^{\mp 2\pi i n x/L} \ \chi(x)$, where
$n=0,1,2,\cdots$, we have
\be
H_{\rm free} = \omega \hbar \sum_{n=1}^{\infty} n^2(\chi^\dagger_{+n}\chi_{+n}+
\chi^\dagger_{-n}\chi_{-n}), \ \ \omega \equiv \frac{2\pi^2 \hbar}{mL^2}.
\label{fermi-free2}
\ee
Note that the zero mode, $\chi_0$, does not enter in the expression
for the hamiltonian. The fermion modes satisfy the canonical
anti-commutation relations,
\be
\{\chi_{\pm n},~\chi^\dagger_{\pm l}\}=\delta_{nl},
\label{anti-commut}
\ee
all other anticommutators vanish.

\subsection{The bosonized hamiltonian}

The bosonization rules that we have developed in Sec. 2 cannot be
applied directly to this hamiltonian because of the degeneracy between
$\pm n$ modes. To get around this difficulty, we will change the
hamiltonian by replacing $n^2$ in the sum in \eq{fermi-free2} by
$(n+\epsilon)^2$, where $\epsilon$ is a small positive real number.
The original problem will be recovered by setting $\epsilon$ to zero
after bosonization is done \footnote{This is, in fact our general rule
for bozonization of a fermionic system described by any other
hamiltonian with degeneracy. The latter is typically due to some
symmetry in the system. One adds extra terms which break all the
symmetries and give a non-degenerate hamiltonian. The parameters of
symmetry breaking are set to zero after completing bosonization. In
this way, one gets a bosonization of the original fermionic
system.}. For a non-zero positive value of epsilon
\footnote{We could alternatively, but completely equivalently,
have chosen a small negative value for epsilon.}, the energy of the
mode for $+n$ is higher than that for $-n$. So the spectrum is now
non-degenerate and looks like that shown in Figure 2. 
 \begin{figure}[htb]
       \centering
       \includegraphics[height=7cm]{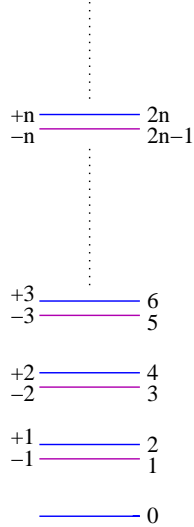}
       \caption{The spectrum with a specified ordering for degenerate levels.}
       \label{fig:2}
     \end{figure}
The next step is to redefine $\chi_0 \equiv \psi_0$, $\chi_{-n}
\equiv\psi_{2n-1}$ and $\chi_{+n} \equiv \psi_{2n}$, $n \neq 0$. The hamiltonian
\eq{fermi-free2} can then be rewritten as
\be
H_{\rm free} = \frac{\omega \hbar}{4} \sum_{n=0}^{\infty} [n+e(n)+(-1)^n 2\epsilon]^2 
\psi^\dagger_n\psi_n.
\label{non-degen-ham}
\ee
Here $e(n)$ vanishes for even $n$ and equals unity for odd $n$, i.e.
\be
e(n) \equiv \frac{1}{2}(1-(-1)^n).
\label{def-e}
\ee
Note that the zero mode piece in \eq{non-degen-ham} vanishes when
$\epsilon$ is set to zero. Since this hamiltonian has the generic form
\eq{fermiham}, its bozonied version can be readily written down using 
\eq{boseham}. Setting $\epsilon$ to zero in the final expression, we get
\be
H_{\rm free} = \frac{\omega \hbar}{4} \sum_{k=1}^N \biggl(\hat n_k+
e(\hat n_k)\biggr)^2, \quad \hat n_k \equiv \sum_{i=k}^N \ad_i a_i +N-k
\label{bose-ham}
\ee
This is the final form of the bosonized hamiltonian for free
non-relativistic fermions on a circle.

\subsection{The relativistic boson approximation}

In his work, Tomonaga showed that for a large number of
non-interacting fermions, small fluctuations of the fermi surface are
described by a free relativistic boson. Our bosonic states diagonalize
$H_{\rm free}$ exactly, even for highly excited states. So it is not
surprising that the relativistic boson is not manifest in our
bosonization. However, we should be able to recover the relativistic
boson from it in the low-energy, low-momentum sector of the large-$N$
limit. In the remainder of this subsection we will explain how that
happens.

The first step is to extract the part of $H_{\rm free}$ which
describes low-energy and low-momentum fluctuations of the two fermi
points. To this end, we note that $H_{\rm free}$ can be rewritten as
follows:
\be
H_{\rm free} = H_F+H_0+H_1,
\label{perturb-ham1}
\ee
where $H_F=\frac{\omega\hbar}{4} \sum_{k=1}^N (N-k+e(N-k))^2$ is the 
energy of the fermi vacuum and 
\bea
H_0 &=& \frac{\hbar \omega N}{2} \biggl(\sum_{k=1}^N k~\ad_k a_k + \hat \nu \biggr),
\label{perturb-ham2}\\
H_1 &=& \frac{\hbar \omega}{4} \sum_{k=1}^N \biggl[\biggl(-k^2+(-1)^N e(k)\biggr)\ad_k a_k
-2\biggl(k(-1)^{N-k}+e(N-k)\biggr)e(\sum_{i=k}^N \ad_i a_i) \nonumber \\
&& ~~~~~~+\biggl(\sum_{i=k}^N \ad_i a_i+(-1)^{N-k} e(\sum_{i=k}^N \ad_i a_i)\biggr)^2\biggr]
\label{perturb-ham3}
\eea
Details of the manipulations required to cast the hamiltonian in this
form suitable for perturbative treatment are given in Appendix
\ref{ham-mani}. In \eq{perturb-ham2}, $\hat \nu=N_--N_{-F}$ is the operator
which measures change in the number of negative momentum fermions in 
the given state
compared to the fermi vacuum. In terms of bosonic occupation number
operators, the number of negative momentum fermions, $N_-$, is given
by the expression
\be
N_- = \sum_{k=1}^N e(\hat n_k),
\label{ferminum}
\ee
where $\hat n_k$ has been defined in (\ref{bose-ham}). Using (\ref{a2}), 
we get
\be
\hat \nu=N_- - N_{-F}
=\sum_{k=1}^N (-1)^{N-k} e(\sum_{i=k}^N \ad_i a_i).
\label{constraint}
\ee

It is clear from the above that for large values of $N$, 
the dominant contribution to the
excitation energy of states in which only a few low energy modes are
present comes from $H_0$, which is a factor of $N$ larger than $H_1$.
In fact, in the large-N limit one can completely ignore the
contribution of $H_1$ on low-energy states. In this sector, then, the
hamiltonian reduces to just $H_0$. This approximation is the
usual linearized approximation to the dispersion relation, which is
valid for low-energy states.

\subsubsection{The partition function}

The simplest way to see that in the large-$N$ limit the hamiltonian
$H_0$ describes a $2$-dimensional massless relativistic boson is to
compute its partition function. Since $\hat \nu$ commutes with $H_0$,
it is natural to classify states by the eigenvalues of this operator,
which we shall denote by $\nu$. We introduce a chemical potential
$\mu$ for it and consider the more general partition function
\be
Z_N(q, y) = \sum_{r_1, r_2, \cdots, r_N=0}^\infty q^{\frac{1}{2}
\sum_{k=1}^N kr_k} 
~y^{\sum_{k=1}^N (-1)^{N-k} e(\sum_{i=k}^N r_i)}, 
\label{partfn}
\ee
where $q=e^{-\hbar \omega N\beta}$ and $y=e^{-\mu}$ and
we have used \eq{constraint}. The
desired partition function, which we shall denote by $Z_N(q)$, 
is obtained by setting $\mu = \frac12\hbar \omega N\beta$,
i.e. $y=q^{1/2}$  in $Z_N(q, y)$.

The partition function $Z_N(q, y)$ satisfies the following recursion
relation:
\be 
Z_N(q, y) = 
(1-q^N)^{-1} [Z_{N-1}(q, y^{-1})+y^{e(N)}q^{N/2} Z_{N-1}(q, y)].
\label{recur}
\ee
A proof of this recursion relation is given in Appendix \ref{rec-rel}. 
Note that the right-hand side is different for even and odd $N$ because
of the appearance of $e(N)$ in this expression. Using
$Z_0(q, y)=1$ in it, one can generate $Z_N(q, y)$ for any value of
$N$. Using mathematica we have checked for a number of values of $N$
that $Z_N(q, y)$ is given by the following analytic expression:
\be
Z_N(q, y) = \sum_{\nu=-\frac{N-1}{2}}^{\frac{N+1}{2}} y^\nu 
q^{\nu(\nu-\frac{1}{2})}
\prod_{n=1}^{\frac{N+1}{2}-\nu}(1-q^n)^{-1} 
\prod_{n=1}^{\frac{N-1}{2}+\nu}(1-q^n)^{-1}.
\label{Z}
\ee
This expression is valid for odd values of $N$ \footnote{For finite
$N$, $Z_N(q)$ has an asymmetry between even and odd $N$. The reason
for this asymmetry is that for odd $N$, the fermi ground state is
unique while for even $N$ the ground state is doubly degenerate. We 
take $N$ to be odd since we want to work with a unique ground state, 
but of course the calculations can just as easily
be done for even values of $N$.}. The upper limit on the sum over $\nu$
is different from the lower limit because there are $(N+1)/2$ fermions
in the fermi vacuum with non-negative momenta, which includes zero
momentum fermion, and in excited states all of them can have negative 
momenta. 

The partition function of a massless relativistic compact boson on a
circle contains a product of two identical factors which come from the 
sum over nonzero oscillator modes \footnote{These are
the oscillator modes of the relativistic boson, not to be confused
with the oscillators $a_k,~\ad_k$ (or $\sigma_k,~\sigmad_k$).  The
oscillator modes of the relativistic boson are more like the
$\rho_{\pm l}$, which can be expressed in terms of
$\sigma_k,~\sigmad_k$, as in equations (\ref{densityp}) and
(\ref{densityn}).}, the contribution from each chiral sector being
$\prod_{n=1}^{\infty} (1-q^n)^{-1}$. In addition, there is a sum
over the eigenvalues of the zero modes in the two chiral sectors,
which can be recast as the lattice sum over momentum and winding
modes. The partition function calculated in (\ref{Z}) has a similar
structure. If we set $y=q^{1/2}$ in it and take a naive large $N$
limit, ignoring the $\nu$ dependence in the product factors, then this
partition function matches precisely with the partition function for a
massless compact scalar in which the zero mode lattice sum is
restricted to momentum modes only. The sum over the winding modes is
missing because of the restriction to a fixed number $N$ of fermions
\footnote{We would like to thank A. Dabholkar and S. Minwalla for a
discussion on this point.}.

It is rather remarkable that in the process of keeping track of $\nu$,
a bunch of harmonic oscillator states is transformed into a massless
relativistic boson. At large but finite $N$, this is only
approximately true, as is evident from the expression in
(\ref{Z}). The corrections go as $e^{-N}$. They have interesting
interpretation in Yang-Mills theory on a cylinder, which is known to be
related to the Tomonaga problem. This is discussed further in Section
6.

\section{Free nonrelativistic fermions on a circle - $1/N$ effects}

In this section we would first like to identify the operators which
create the single-particle states of a massless relativistic boson,
which we have counted in the calculation of the partition function
above. We would then like to incorporate the effects of a small but
non-zero value of $1/N$ on this free relativistic boson. In the
following discussion we will assume that the large-$N$ limit is taken
through odd values.

\subsection{States}

The operator $\hat \nu$ commutes with both $H_0$ and $H_1$ separately, as
can be easily verified. The label $\nu$ on the states of the massless
relativistic boson is therefore a conserved quantum number. Setting
$y=q^{1/2}$ in the expression for the partition function given in
(\ref{Z}), one finds that the lowest amount of energy carried by a
state labeled by the value $\nu$ is $\hbar \omega N \nu^2$. The
states which realize these eigenvalues for the operator $\hat \nu$ are
easily constructed by inspection of the fermionic states. Using the
obvious notation $\ket{\nu}$ for them, we have 
\be
\ket{\nu} = \left\{
\begin{array}{l}
\sigmad_{2\nu-1}\sigmad_{2\nu-2} \cdots \sigmad_1 \ket{0}, 
\quad \nu > 0 \\ ~ \\
\sigmad_{2|\nu|}\sigmad_{2|\nu|-1} \cdots \sigmad_1 \ket{0}, 
\quad \nu < 0.
\end{array}~~~~ \right.
\label{nu}
\ee
The following properties can be easily verified:
\be
\hat \nu \ket{\nu} = \nu \ket{\nu}, \quad H_0 \ket{\nu} = 
\hbar \omega N \nu^2 \ket{\nu}, \quad H_1 \ket{\nu} = 0.
\label{nu2}
\ee
Since $\ket{\nu}$ is the lowest energy state in the sector labeled by
$\nu$, it acts as a sort of ``vacuum'' state in that sector. A tower
of excited states can then be created by the oscillator modes of the
massless boson on each of these vacuua.  In order to keep the
following discussion simple, we will restrict ourselves to states in
the $\nu=0$ sector. The discussion can be easily generalized to
arbitrary values of $\nu$. At the end we will indicate the changes
that need to be made to accommodate general $\nu$.

In the $\nu=0$ sector, at the lowest excitation level there are two
states, $$(\sigmad_1)^2 \ket{0}, \ \sigmad_2 \ket{0}.$$ Using the
rules of bosonization discussed in Sec 2, their excitation momentum
can be seen to be the lowest possible and of opposite sign. These are
the expected two single-particle chiral states at lowest
energy. Moreover, since they are also eigenstates of the full
hamiltonian $H_{\rm free}$, one can calculate their $H_1$
eigenvalues. For odd $N$ these vanish, as can be easily checked using
\eq{perturb-ham3}. The absence of a ``tree-level'' $1/N$ correction
(which is the same as the correction in first-order perturbation
theory) to the energy of these states is consistent with their
interpretation as one-particle states of a massless relativistic
boson.

At the next level of excitation, we have five possible states:
$$(\sigmad_1)^4 \ket{0},~\sigmad_1\sigmad_3
\ket{0},~(\sigmad_1)^2\sigmad_2 \ket{0},~\sigmad_4 \ket{0},
~(\sigmad_2)^2 \ket{0}.$$ It is easy to see that the last state has
the same total momentum as the fermi vacuum, so it accounts for the
non-chiral state with two single-particle lowest states of opposite
chirality expected at this level of excitation. The other four states
must, therefore, account for the expected two single-particle chiral
states and the two two-particle chiral states obtained from the lowest
single-particle chiral states. As at the lowest excitation level,
these states are eigenstates of the full hamiltonian, and therefore
also that of $H_1$. However, unlike in the above case, the eigenvalues
do not all vanish. This is not necessarily a problem since consistency
of interpretation as states of a massless relativistic boson is
ensured if we can form their linear combinations which are such that
the expectation value of $H_1$ vanishes in all of them. 
Such linear combinations can, in fact, be formed.  These are:
$$\frac{1}{\sqrt 2}[(\sigmad_1)^4 \pm
\sigmad_1\sigmad_3]\ket{0}; \  \frac{1}{\sqrt
2}[(\sigmad_1)^2\sigmad_2 \pm \sigmad_4]\ket{0},$$ with opposite sign
of momentum for each of the second pair of states compared to the
first pair. In each of these orthogonal pairs, we would like to
interpret one of the linear combinations as a single-particle state of
the massless boson, while the orthogonal combination would be a
two-particle state of the same momentum and energy. Also, $H_1$ sends
a linear combination to its orthogonal combination. This is why its
expectation value in any of these states vanishes. But this also means
that $H_1$ has non-zero matrix elements between the two states of the
orthogonal combinations. These matrix elements of $H_1$, which connect
a single-particle state with a two-particle state, can be interpreted
as low-energy scattering amplitudes in a relativistic field theory of
a massless scalar with a cubic coupling whose strength goes as $1/N$.
Finally, also note that the remaining state $(\sigmad_2)^2 \ket{0}$
with zero net momentum has a vanishing $H_1$ eigenvalue.

The above analysis can be extended to higher excited states in a
straightforward manner. The results are similar; for odd $N$ one can
always find orthogonal linear combinations which are such that the
expectation value of $H_1$ vanishes in these combinations, while the
matrix elements are in general non-zero.

In Tomonaga's work, the modes of the relativistic boson are related to
modes of spatial fermion density operator. We should, therefore,
expect a relation between the linear combinations we have found above
and the modes of fermion density operator.  It turns out that for
low-energy excitations, precisely one of the linear combinations in
each of the two chiral sectors is an appropriate mode of the fermion
density:
\bea
\frac{1}{\sqrt l} \sum_{n={\rm even}} \psid_{n+2l} \psi_n \ket{F_0} \equiv 
\rho^\dagger_{+l} \ket{0}
 &=& \frac{1}{\sqrt l} \sum_{k=1}^l (\sigmad_1)^{2(l-k)+1} \sigmad_{2k-1} 
\ket{0}, 
\label{densityp}\\
\frac{1}{\sqrt l} \sum_{n={\rm odd}} \psid_{n+2l} \psi_n \ket{F_0} \equiv 
\rho^\dagger_{-l} \ket{0} 
&=& \frac{1}{\sqrt l} \sum_{k=1}^l (\sigmad_1)^{2(l-k)} \sigmad_{2k} \ket{0}, 
\label{densityn}
\eea
where $\ket{F_0}$ is the fermi ground state, which is in fact also the
bosonic ground state, $\ket{0}$. These expressions are valid only for
$l \leq \frac{N+1}{2}$ \footnote{For $l > \frac{N+1}{2}$, the sum over
$k$ terminates at $k=\frac{N+1}{2}$, irrespective of the actual value
of $l$. Since there cannot be any single-particle states for $l >
\frac{N+1}{2}$, these states must be interpreted as multi-particle states.}, 
with $N$ having been assumed to be odd. Moreover, the fermion
bilinears in \eq{densityp}, \eq{densityn}, which are equivalent to the
chiral bilinears, $\sum \chi_{\pm (n+l)}^\dagger
\chi_{\pm n}$, are related to the modes of the original fermion
density only for sufficiently small values of $l$
\footnote{Very high energy modes of the fermion density also involve
mixed chirality fermion bilinears. This is discussed in detail in
Section 5.} such that there are no holes deep inside the fermi
sea. The examples considered above correspond to $l=1,~2$.

These fermion bilinears may be interpreted as single-particle states
of the massless relativistic boson at energy $l$ and momentum $\pm
l$. Multiple applications of the fermion bilinears on the fermi vacuum
must then, for consistency, reproduce all the other linear
combinations at any excitation energy and momentum level. It is easy
to check that this is true for a few low lying levels. Consider, for
example, the state $(\rho^\dagger_{+1})^2
\ket{0}$. Using \eq{densityp} and the bosonization rules of Sec 2, 
it is easy to see that 
\be
(\rho^\dagger_{+1})^2 \ket{0} = \rho^\dagger_{+1} (\sigmad_1)^2 \ket{0}
= [(\sigmad_1)^4 - \sigmad_1\sigmad_3] \ket{0}. 
\label{multipart}
\ee
The minus sign in the second term on the right-hand side above comes
from a fermion annihilation operator crossing over a fermion in the
vacuum state. This linear combination of oscillator states is
orthogonal to the combination that appears in $\rho^\dagger_{+2}
\ket{0}$. Some other examples of small $l$ chiral states are:
\be
(\rho^\dagger_{+1})^3 \ket{0}=[(\sigmad_1)^6-2(\sigmad_1)^3 \sigmad_3
+\sigmad_1 \sigmad_5] \ket{0}, \ \rho^\dagger_{+1} \rho^\dagger_{+2} \ket{0}=
\frac{1}{\sqrt 2}[(\sigmad_1)^6-\sigmad_1 \sigmad_5] \ket{0}.
\label{multipart2} 
\ee
An example of a non-chiral multi-particle state is
\be
\rho^\dagger_{+1} \rho^\dagger_{-1} \ket{0}=(\sigmad_2)^2 \ket{0}.
\ee
Both the states in \eq{multipart2} are orthogonal to the
single-particle state for $l=3$. This holds true in a few other
small-$l$ examples that we have checked. We believe that it is
generally true for $l \leq \frac{N+1}{2}$.  It would be nice to
have a general proof of this statement.

\subsection{Interactions}

In the large-$N$ and low-energy limit, non-interacting
non-relativistic fermions in one space dimension are known to be
described by a collective field theory \cite{JS} of a
massless boson with a cubic coupling which is of order $1/N$. We have
already mentioned possible O($1/N$) interactions among the density
modes $\rho_{\pm l}$ in the previous subsection. Here we will discuss
these interactions in more detail. As we will see, a cubic interacting
boson theory arises in the large-$N$ and low-energy sector of our
bosonized theory, like in the collective field theory approach. The
difference is that we have a greater and more systematic control on
$1/N$ corrections. We can of course also go beyond the low-energy
large-$N$ approximation, where the local cubic scalar field theory
description breaks down, since our bosonization is exact.

Consider the action of $H_1$, \eq{perturb-ham3}, on the states
$\rho^\dagger_{\pm l} \ket{0}$. Each of the oscillator states occuring
in the sum in \eq{densityp}, \eq{densityn} is an eigenstate of $H_1$,
with an eigenvalue that can be easily computed. In fact, 
\be
H_1 (\sigmad_1)^r \sigmad_s \ket{0} = 
\left\{ \begin{array}{l}
\frac{\hbar \omega}{4} (r^2-s^2) (\sigmad_1)^r \sigmad_s \ket{0},
\ \ \ r, \ s \ {\rm odd,~including~} s=1, \\ ~ \\
\frac{\hbar \omega}{4}((r+1)^2-(s-1)^2) (\sigmad_1)^r \sigmad_s \ket{0}
\ \ \ r, \ s \ {\rm even,~} s \neq 0.
\end{array}~~~~ \right.
\ee
It follows that
\bea
H_1 \rho^\dagger_{+l} \ket{0} &=&  \frac{\hbar \omega}{\sqrt l} \sum_{k=1}^l 
[l(l+1)-2lk] (\sigmad_1)^{2(l-k)+1} \sigmad_{2k-1} \ket{0},
\label{h1} \\
H_1 \rho^\dagger_{-l} \ket{0} &=&  \frac{\hbar \omega}{\sqrt l} \sum_{k=1}^l 
[l(l+1)-2lk] (\sigmad_1)^{2(l-k)} \sigmad_{2k} \ket{0}.
\label{h2}
\eea
It is now easy to verify that for $l \leq \frac{N+1}{2}$,
$\bra{0}\rho_{\pm l} H_1 \rho^\dagger_{\pm l} \ket{0}=0$ 
\footnote{For $l > \frac{N+1}{2}$, it can be easily checked that $H_1
\rho_{\pm}^\dagger \ket{0}$ is not orthogonal to the state 
$\rho_{\pm}^\dagger \ket{0}$. Remember that for $l > \frac{N+1}{2}$,
the sum over $k$ in (\ref{densityp}) and (\ref{densityn}) has to be
truncated at $\frac{N+1}{2}$.}, so that the linear combinations of the
oscillator states which appear in \eq{h1} and \eq{h2} must be
multi-particle states of the massless boson. We have already seen
examples of such linear combinations in \eq{multipart} and
\eq{multipart2}. These are special cases of \eq{h1} for $l=2,~3$.

Notice that a specific linear combination of oscillator states enters
on the right-hand side of \eq{h1} (same is true of \eq{h2}). For
example, for $l=3$ there are two different multi-particle states in
each chiral sector, $\rho^\dagger_{\pm 1} \rho^\dagger_{\pm 2}
\ket{0}$ and $(\rho^\dagger_{\pm 1})^3 \ket{0}$, but only the former
enters in \eq{h1} and \eq{h2}, which is a two-particle state. For
$l=4$ we have a more non-trivial example. Here, one can show that
\be
H_1 \rho^\dagger_{+4} \ket{0}=4 \hbar \omega [{\sqrt 3} \rho^\dagger_{+1} 
\rho^\dagger_{+3} \ket{0}+(\rho^\dagger_{+2})^2] \ket{0}.
\ee
So in this case the right-hand side is a linear sum of the two
possible two-particle states. More generally, for $l \leq
\frac{N+1}{2}$, $H_1\rho^\dagger_{+l}
\ket{0}$ can be written as a linear sum of all possible 
two-particle states only:
\be
H_1\rho^\dagger_{\pm l} \ket{0}=\hbar \omega \sum_{m=1}^{l-1} c^{l}_{\pm m} 
\rho^\dagger_{\pm m}
\rho^\dagger_{\pm(l-m)} \ket{0}, \ \ l \leq \frac{N+1}{2}.
\label{coeffc}
\ee
We have calculated the coefficients $c^{l}_{\pm m}$. The calculation
is described in Appendix \ref{calcu}. We get,
\be
c^{l}_{\pm m}=\sqrt{lm(l-m)}.
\label{c}
\ee

The above calculations can be summarized as follows. In a field theory
setting, the low-energy and large-$N$ limit of the fermi system under
discussion is described by the following hamiltonian of a massless 
relativistic scalar in $2$-dimensions with a cubic coupling:
\bea
H &=& \hbar \omega \bigg[N \sum_l l (\phi_{+l}^\dagger \phi_{+l}+
\phi_{-l}^\dagger \phi_{-l}) \nonumber \\
&&~~~~+\biggl\{\biggl(
\sum_{l} \sum_{m=1}^{l-1} \sqrt{lm(l-m)} \phi_{+(l-m)}^\dagger 
\phi^\dagger_{+m} \phi_{+l}+{\rm c.c.}\biggr)+(+ \rightarrow -)
\biggr\} \bigg].
\eea
This is an effective hamiltonian which is valid only for energies $l
<< N$ since it is only for low energies that $\phi$'s satisfy the
approximate commutation relation, $[\phi_{\pm l},
\phi_{\pm m}^\dagger] \sim \delta_{lm}$. It describes the effective 
low-energy dynamics in the $\nu=0$ sector. 

It is straightforward to extend the above discussion to states with
arbitrary $\nu$. The single-particle states of the massless boson in
the two chiral sectors can be obtained by applying the operators
$\rho_{\pm l}^\dagger$ on the corresponding ``vacuum'' state,
$\ket{\nu}$, similar to equations (\ref{densityp}) and
(\ref{densityn}) which give the states in the $\nu=0$ sector. We have
already seen in equation (\ref{nu2}) that for $\nu \neq 0$ there is an
additional O$(N)$ term, $\hbar \omega N \nu^2$, in the effective
hamiltonian. It turns out that there is an additional term in the 
effective low-energy hamiltonian, but this term is O$(1)$ in $N$. 
It arises because the expectation value of $H_1$ does not vanish in the 
states with a nonzero value of $\nu$. The complete effective low-energy 
hamiltonian turns out to be
\bea
H &=& \hbar \omega \bigg[N \nu^2+N \sum_l l (\phi_{+l}^\dagger \phi_{+l}
+\phi_{-l}^\dagger \phi_{-l})-2 \nu \sum_l l (\phi_{+l}^\dagger \phi_{+l}
-\phi_{-l}^\dagger \phi_{-l}) \nonumber \\
&&~~~~+\biggl\{\biggl(
\sum_{l} \sum_{m=1}^{l-1} \sqrt{lm(l-m)} \phi_{+(l-m)}^\dagger 
\phi^\dagger_{+m} \phi_{+l}+{\rm c.c.}\biggr)+(+ \rightarrow -)
\biggr\} \bigg].
\eea
A cubic effective hamiltonian for nonrelativistic fermions in one
space dimension was first obtained in the collective theory approach
by \cite{JS}. Our expression for the hamiltonian agrees
with that obtained by \cite{LMR} (see also \cite{MP}) in the context 
of two-dimensional Yang-Mills theory on a circle.

\section{Beyond low-energy effective theory}

As we have seen in the previous sections, the cubic bosonic low-energy
effective field theory is a result of a controlled
large-$N$ and low-energy limit of a more complete finite-$N$
bosonization of the fermi system. In this more complete setting it is
naturally possible to go beyond the low-energy approximation and to do
calculations, for example of the correlation functions, at high
energies and finite $N$. At high energies, there are two distinct ways
in which we must modify the perturbative calculations of the previous
section. Firstly, at high energies the contribution of $H_1$ to the
hamiltonian can be comparable to that of $H_0$, so perturbation
expansion breaks down. In the non-interacting theory this can be
taken care of by doing exact calculations. Secondly, at high
energies the states created from the vacuum by the modes of the
fermion density operator, $\tilde \rho_{\pm l} \equiv
\frac{1}{\sqrt{l}}\int dx~e^{\mp 2\pi i l
x/L}~\chi^\dagger(x)\chi(x)$, are not identical to the single-particle
states created by $\rho_{\pm l}$ from the vacuum. In fact, we have
\bea
\tilde \rho^\dagger_{+l} \ket{0} &=& \rho^\dagger_{+l} \ket{0}+
\frac{1}{\sqrt{l}} \sum_{n=1}^{l} \psid_{2(l-n)} \psi_{2n-1} \ket{0}, 
\nonumber \\
\tilde \rho^\dagger_{-l} \ket{0} &=& \rho^\dagger_{-l} \ket{0}+
\frac{1}{\sqrt{l}}\sum_{n=1}^{l} \psid_{2n-1} \psi_{2(l-n)} \ket{0}. 
\label{rho-t-states}
\eea
In the above equations, the second term on the right-hand side
contributes only for $l > \frac{N+1}{2}$. It is a $\nu=-1$
state. Such terms will show up as extra contributions in scattering
amplitudes. Consider, for example, the ``two-particle'' state $\tilde
\rho^\dagger_{+m} \tilde \rho^\dagger_{+(l-m)} \ket{0}$. For $l \geq
\frac{N+3}{2}$, this state is not the same as the state
$\rho^\dagger_{+m} \rho^\dagger_{+(l-m)} \ket{0}$, but has an extra
contribution. For example, let us take $l=\frac{N+3}{2}$. Then,
\be
\tilde \rho^\dagger_{+m} \tilde \rho^\dagger_{+(\frac{N+3}{2}-m)} \ket{0}=
\rho^\dagger_{+m} \rho^\dagger_{+(\frac{N+3}{2}-m)} \ket{0}+
\frac{1}{\sqrt{m(\frac{N+3}{2}-m)}} \sigmad_1 \sigmad_{N-1} \ket{0}.
\ee 
The extra term on the right-hand side above will contribute in
the calculation of the 3-point function, $\bra{0} \tilde
\rho^\dagger_{+\frac{N+3}{2}} \tilde \rho^\dagger_{+m} \tilde
\rho^\dagger_{+(\frac{N+3}{2}-m)} \ket{0}$, because of the extra term
in (\ref{rho-t-states}).

Basically, the point is that at low energies, it is possible to
restrict the states created by the modes of the fermion density to the
$\nu=0$ sector. However, at high energies, effects of the $\nu
\neq 0$ states will show up, under appropriate conditions, in the
correlation functions of the modes of the fermion density. 

\section{Interacting non-relativistic fermions on a circle}

The full fermionic hamiltonian is a sum of the free part and interactions,
\be
H = H_{\rm free}+H_{\rm int}
\label{fermi-ham}
\ee
Following Tomonaga, we will assume an interaction of the
following form (which could arise for example from the Coulomb force
between the fermions):
\be
H_{\rm int} = \int_0^L dx \int_0^L dy ~\tilde \rho(x) \tilde \rho(y) J(x-y), 
\label{fermi-int1}
\ee
where $\tilde \rho(x)=\chi^\dagger(x)\chi(x)$ is the fermion spatial
density and $J(x-y)$ is the fermion-fermion interaction potential. If
the interaction has a range much larger than $L/N$, it is a good
approximation to replace $H_{\rm int}$ by $\sum_l l(J_{+l}
\phi_{+l}^\dagger \phi_{+l}+J_{-l}\phi_{-l}^\dagger \phi_{-l})$, 
where $J_{\pm l}$ are the fourier modes of the interaction potential;
the sum has a cut-off at some $l << N$ because of the long range of
the interaction potential and so it is consistent to restrict to
states with $\nu=0$ . For potentials with a range smaller than $L/N$,
one must take into account the fact that the modes of the fermion
density have extra terms like that in the equation
(\ref{rho-t-states}). In this case, interactions can excite states
with $\nu \neq 0$ and one needs to take these into account in
calculations.

\section{Two-dimensional Yang-Mills on a cylinder}

Tomonaga's problem has surprising connections with a variety of
interesting problems in field theory and string theory. In fact, it is
known for some time now that non-relativistic fermions appear in
two-dimensional Yang-Mills on a cylinder with U$(N)$ gauge group
\cite{MP}. In recent works it has been pointed out that they also appear 
in the physics of black holes \cite{V,DGOV}, with possible
connections to the physics of baby universe creation. Because of this,
our bosonization has applications to these problems as well. Below we
will briefly elaborate on these connections.

Two-dimensional Yang-Mills theory on a cylinder can be shown to be
equivalent to a string theory \cite{G,GT}. See also
\cite{Wadia,CMR1,CMR2,MD}. In this context, an interesting observation
due to \cite{LMR} is that states with excitation energy of O$(N^2)$
are D1-branes. This observation is mainly based on the presence of
O$(e^{-1/g_s} \sim e^{-N})$ contributions to the partition function
\cite{LMR,KJR}. As discussed below, such contributions are also 
present in (\ref{Z}) for large finite $N$.

For connection with black holes and baby universes, one considers Type
IIA string theory on CY supporting a supersymmetric configuration
of D4, D2 and D0 branes.  The back-reacted geometry is a black hole in
the remaining four non-compact directions, which is characterized by
the D4, D2 and D0 charges. It was shown in \cite{V} that the bound state of
D4, D2 and D0 branes maps to the partition function of pure two-dimensional
Yang-Mills theory on a cylinder. The point of \cite{DGOV}, relevant
for the present discussion, is that the partition function with a
given asymptotic charge must necessarily include multi-centered black
holes, corresponding to configurations with multiple filled bands of
fermion energy levels. In the black hole context, the near-horizon
limit of the multi-centred configurations gives rise to an ensemble of
$AdS_2 \times S^2$ configrations. The existence of such multiple
configrations gives rise to nonperturbative corrections to the OSV
\cite{OSV} relation; schematically
\be
Z_{BH} = | \psi|^2 + O(e^{-N})
\label{osv1}
\ee
The uncorrected equation is valid for a single black hole, and
corresponds in the fermion theory to two decoupled fermi surfaces (at
the top and at the bottom) which is the correct description in the $N
\rightarrow \infty$ limit. The O$(e^{-N})$ corrections arise in the
black hole context from the fact that partition function over
geometries with a given asymptotic charge $Q$ includes multiple black
holes with charges $Q_i$ such that $Q=\sum Q_i$. The contribution of
these to the total partition function is given by $e^{(-I)},~I = \sum
S_{BH}(Q_i)$. In case of the fermion theory, the $O(e^{-N})$
corrections signify the fact that the at finite $N$, the approximation
of the Fermi sea as having two infintely separated Fermi surfaces is
not valid and includes in the partition function many more states than
actually exist in the system; the corrections subtract those states
iteratively.

In our present formalism, the structure of equation (\ref{osv1}) can be
recognized in (\ref{Z}). Setting $y=q^{1/2}$ in it, we get
\be
Z_N(q) = \sum_{\nu=-\frac{N-1}{2}}^{\frac{N+1}{2}} q^{\nu^2}
\prod_{n=1}^{\frac{N+1}{2}-\nu}(1-q^n)^{-1} 
\prod_{n=1}^{\frac{N-1}{2}+\nu}(1-q^n)^{-1}.
\label{Z2}
\ee
Writing out the first few terms in the $\nu$ sum explicitly, we get
\bea
Z_N(q) &=& \biggl(1+q+q \frac{1-q^{\frac{N-1}{2}}}{1-q^{\frac{N+1}{2}+1}} 
+ \cdots \biggr)
\prod_{n=1}^{\frac{N+1}{2}+1}(1-q^n)^{-1} 
\prod_{n=1}^{\frac{N-1}{2}}(1-q^n)^{-1} \nonumber \\
&=& \biggl(1+2q-q^{\frac{N+1}{2}}+ \cdots \biggr)
\prod_{n=1}^{\frac{N+1}{2}}(1-q^n)^{-1} 
\prod_{n=1}^{\frac{N-1}{2}}(1-q^n)^{-1} \nonumber \\
&=& \biggl(1+2q-q^{\frac{N+1}{2}}+ \cdots \biggr)
\biggl(1-q^{\frac{N+1}{2}}+ \cdots \biggr) 
\bigg[\prod_{n=1}^\infty (1-q^n)^{-1}\bigg]^2. 
\label{Z3}
\eea
We see that there are two types \cite{DGOV} of O$(e^{-N})$
corrections. One arises from the first factor which originated from
sum over $\nu$. The other arises from the second factor which came
from writing the two product factors (in the second equality above) as
their $N \rightarrow \infty$ limit and the deficit. This division is
of course arbitrary, only the overall structure of the final result,
which is as indicated in (\ref{osv1}), being meaningful. We see that
the hamiltonian (\ref{perturb-ham2}) of a bunch of harmonic
oscillators provides a simple example of the two types of
nonperturbative corrections discussed in \cite{DGOV}.

\section{\label{discussion}Summary and discussion}

In this paper we have used the tools recently developed by us
\cite{DMS} for an
exact bosonization of a finite number $N$ of non-relativistic fermions
to discuss the classic Tomonaga problem. We have shown that the
standard cubic effective hamiltonian for a massless relativistic boson
arises in a systematic large-$N$ and low-energy limit. At finite $N$
and high energies, however, the low-energy effective description
breaks down and the exact bosonized hamiltonian must be used. A
curious feature of this exact bosonized theory is that there is no
underlying space visible. The latter emerges only in the semiclassical
(large-$N$) limit at low energies. Our bosonized theory thus provides
an interesting example of this phenomenon which is expected to be a
generic property of any consistent theory of quantum gravity. 

Tomonaga's problem has an interesting connection with pure Yang-Mills
theory on a cylinder. In the context of the recent discussion of baby
universes in string theory black holes, we have pointed out that the
O$(N)$ piece in our bosonized hamiltonian provides a simple model
for understanding the origin of two different kinds of nonperturbative
O$(e^{-N})$ corrections to the partition function. We may recall here
that in the application of our bosonization to the half-BPS sector of
${\cal N}=4$ super Yang-Mills theory \cite{DMS,DMM}, 
our bosonic oscillators turned
out to create single-particle giant graviton states from the
\ads~ground state. It would be intersting to investigate whether our
bosonic oscillators also have a natural interpretation in the baby
universe context.

It is possible to generalize our bosonization to higher space
dimensions. A brief discussion of this has been given in Appendix D. An
interesting aspect of the bosonized theory seems to be the absence of
a manifest reference to the number of dimensions. This is not really
surprising since space-time emerges only in the semiclassical
low-energy limit in our bosonized theory. It would be interesting to
further explore the bosonized theory in higher dimensions. In
particular, it would be interesting to see how the bosonized theory
encodes symmetries, e.g. spatial rotations.

\newpage

\appendix

\section{\label{ham-mani} Calculation of perturbative form of hamiltonian}

In this appendix we will give details of the derivation of
(\ref{perturb-ham1})-(\ref{perturb-ham3}) from (\ref{bose-ham}). The
first step is to rewrite (\ref{bose-ham}) as follows:
\be
H_{\rm free} = \frac{\hbar \omega}{4} \sum_{k=1}^N \biggl([N-k+e(N-k)]+
[\sum_{i=k}^N \ad_i a_i+(-1)^{N-k} e(\sum_{i=k}^N \ad_i a_i)]\biggr)^2.
\ee
In writing this, we have made use of the identity 
\be
e(\sum_{i=k}^N
\ad_i a_i +N-k)=e(N-k)+(-1)^{N-k}e(\sum_{i=k}^N \ad_i a_i),
\label{a2}
\ee
which can be easily derived from the definition of $e(n)$ given in
(\ref{def-e}). Opening up the square, the first term in the square
brackets gives rise to $H_F$. The cross term is 
\be
\frac{\hbar \omega}{2} \sum_{k=1}^N [N-k+e(N-k)][\sum_{i=k}^N 
\ad_i a_i+(-1)^{N-k} e(\sum_{i=k}^N \ad_i a_i)].
\ee
Using 
\be
\sum_{k=1}^N g(k) \sum_{i=k}^N \ad_i a_i=\sum_{k=1}^N
f(k) \ad_k a_k, \ f(k)=\sum_{i=1}^k g(i),
\ee
the cross term can be rewritten as 
\be
\frac{\hbar \omega}{2} \sum_{k=1}^N [kN-\frac{k^2}{2}+\frac{(-1)^N}{2} e(k)]\ad_k a_k+
\frac{\hbar \omega}{2} \sum_{k=1}^N [N-k+e(N-k)](-1)^{N-k} e(\sum_{i=k}^N \ad_i a_i).
\ee
The leading term is $\frac{\hbar \omega N}{2} \sum_{k=1}^N [k\ad_k a_k+(-1)^{N-k}
e(\sum_{i=k}^N \ad_i a_i)]$, which is just $H_0$. The rest combines
with the square of the second term in $H_{\rm free}$ above to give
$H_1$.

\section{\label{rec-rel} Recursion relation for the partition function}

In this Appendix we will give details of the derivation of the
recursion relation (\ref{recur}). First we note that
\be
e(\sum_{i=k}^{N} r_i)=\left\{
\begin{array}{l}
e(\sum_{i=k}^{N-1} r_i), \ r_N \ {\rm even}, \\ ~ \\
1-e(\sum_{i=k}^{N-1} r_i), \ r_N \ {\rm odd},
\end{array}~~~~ \right.
\ee
from which it follows that 
\bea
\sum_{k=1}^N (-1)^{N-k} e(\sum_{i=k}^N r_i) &=& e(r_N)+
\sum_{k=1}^{N-1} (-1)^{N-k} e(\sum_{i=k}^N r_i) \nonumber \\
&=& \left\{
\begin{array}{l}
-\sum_{k=1}^{N-1} (-1)^{N-1-k} e(\sum_{i=k}^{N-1} r_i), \ r_N \ {\rm even}, \\ ~ \\
e(N)+\sum_{k=1}^{N-1} (-1)^{N-1-k} e(\sum_{i=k}^{N-1} r_i), \ r_N \ {\rm odd}.
\end{array}~~~~ \right.
\eea
Using this result in (\ref{partfn},) we can explicitly do the summation
over $r_N$ by writing it as separate sums over even and odd $r_N$. We get,
\bea
Z_N(q, y) &=& \sum_{r_N={\rm even}}  q^{\frac{1}{2} Nr_N} 
\sum_{r_1, r_2, \cdots, r_{N-1}=0}^\infty q^{\frac{1}{2}\sum_{k=1}^{N-1} kr_k} 
~y^{-\sum_{k=1}^{N-1} (-1)^{N-1-k} e(\sum_{i=k}^{N-1} r_i)} \nonumber \\ &+&
y^{e(N)} \sum_{r_N={\rm odd}} q^{\frac{1}{2} Nr_N}
\sum_{r_1, r_2, \cdots, r_{N-1}=0}^\infty q^{\frac{1}{2}\sum_{k=1}^{N-1} kr_k} 
~y^{\sum_{k=1}^{N-1} (-1)^{N-1-k} e(\sum_{i=k}^{N-1} r_i)}. \nonumber \\ 
\eea
Equation (\ref{recur}) now trivially follows from this.

\section{\label{calcu} Calculation of the coefficients $c^l_{\pm m}$}

In this Appendix, we will give details of the calculation of the
coefficients $c^l_{\pm m}$ which determine the tree level scattering
amplitudes through the equation (\ref{coeffc}). For definiteness, the
following calculations are done for the '$+$' sign for $l$
odd. Calculations for the other choices can be done similarly.

We need to prove the following identity:
\be
\sum_{m=1}^{l-1} \sqrt{m(l-m)} \rho^\dagger_{+m} \rho^\dagger_{+(l-m)} \ket{0}=
\sum_{k=1}^l [l(l+1)-2lk] (\sigmad_1)^{2(l-k)+1} \sigmad_{2k-1} \ket{0}.
\ee
For $l$ odd, this can be rewritten as 
\be
\sum_{m=1}^{\frac{l-1}{2}} \sqrt{m(l-m)} \rho^\dagger_{+m} \rho^\dagger_{+(l-m)} \ket{0}=
\sum_{k=1}^l [\frac{l+1}{2}-k] (\sigmad_1)^{2(l-k)+1} \sigmad_{2k-1} \ket{0}.
\label{identity}
\ee
In terms of the fermion bilinears, the left-hand side of this identity is
\be
\sum_{m=1}^{\frac{l-1}{2}} \sum_{n={\rm even}} \psid_{n+2m} \psi_n 
\sum_{p={\rm even}} \psid_{p+2(l-m)} \psi_p \ket{F_0}.
\ee
Figure 3 shows the fermion state one gets after the first fermion
biliner has created a particle-hole pair in the fermi vacuum.
\begin{figure}[htb] 
\centering 
\includegraphics[height=7cm,
width=5cm]{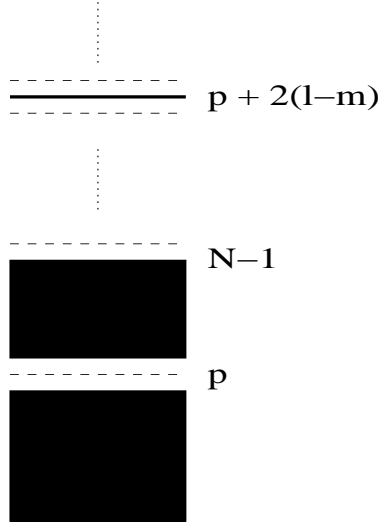} 
\caption{A particle-hole pair state created from vacuum by a single fermion bilinear .}  
\label{fig:3}
\end{figure} 
After the second fermion bilinear has acted on this
state, two types of states can result. One can have a
state with a single particle-hole pair (the second bilinear either
moves the particle up or the hole down), as shown in Figures 4a and
4b.  
\begin{figure}[htb] 
\centering 
\includegraphics[height=7cm, width=4cm]{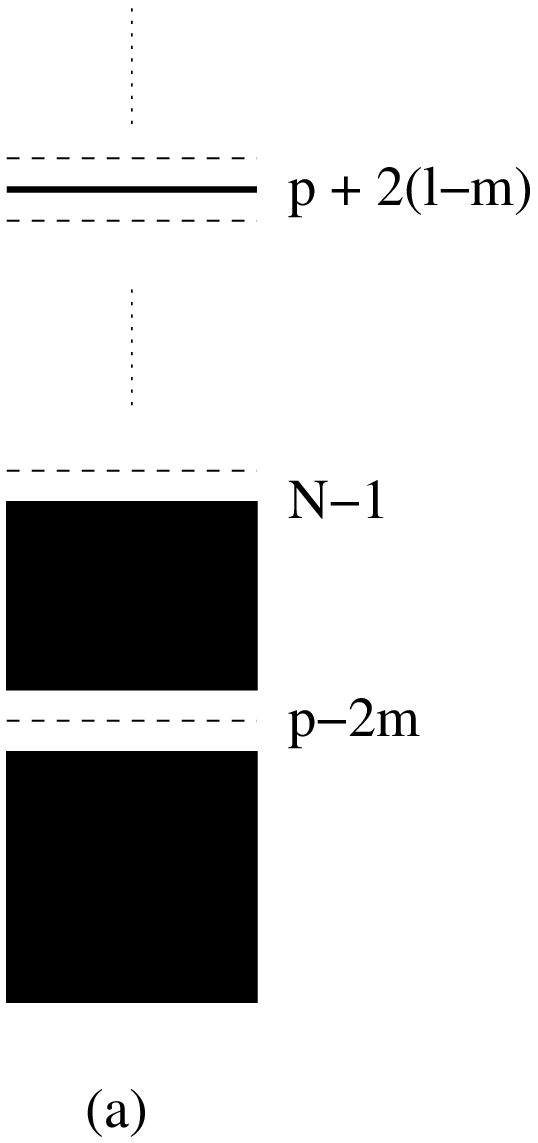} 
\hspace{3ex} 
\includegraphics[height=7cm, width=4cm]{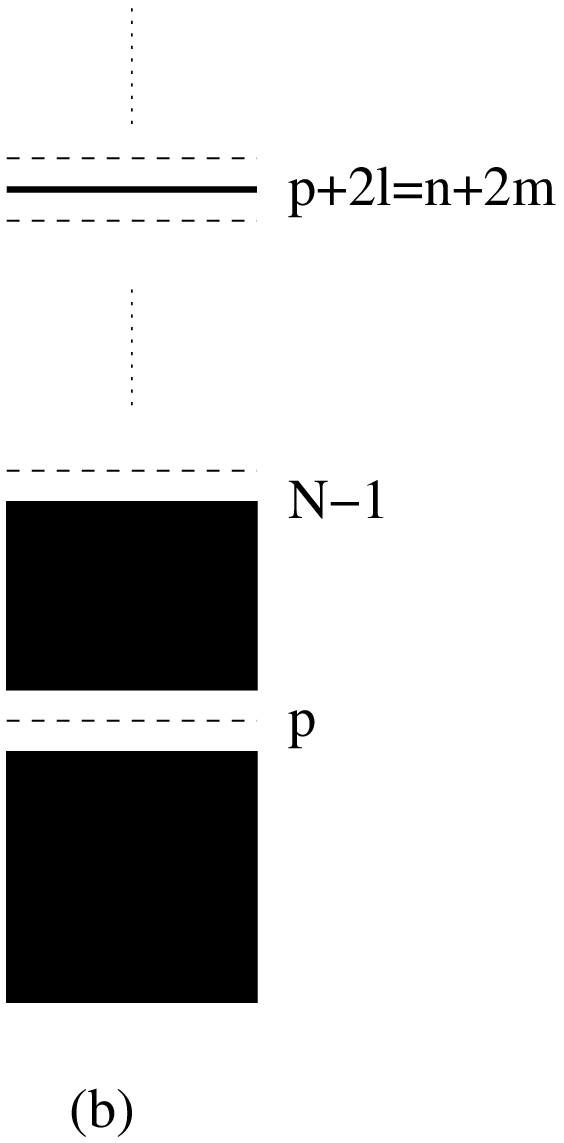} 
\caption{The state with a single particle-hole pair resulting from the action of two fermion bilinears on vacuum .}  
\label{fig:4}
\end{figure} 
Or one can have a state with two particle-hole pairs, as
shown in Figures 5a and 5b.  
\begin{figure}[htb] 
\centering
\includegraphics[height=7cm, width=3.5cm]{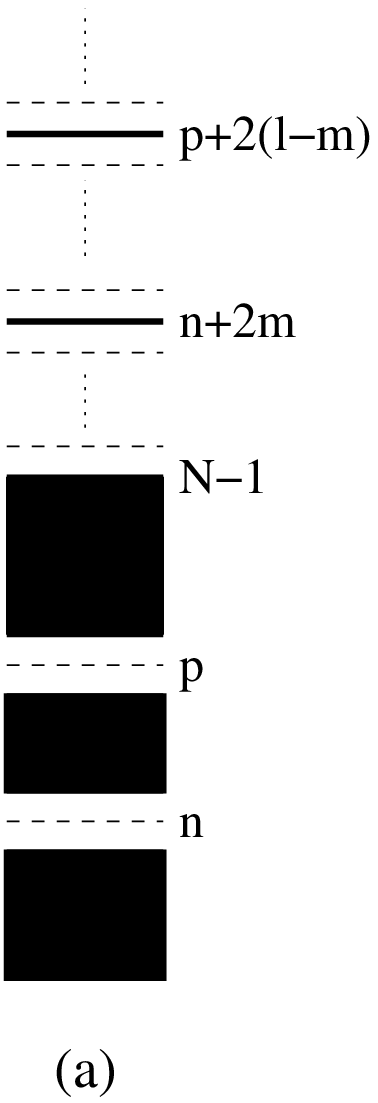} 
\hspace{3ex}
\includegraphics[height=7cm, width=3.5cm]{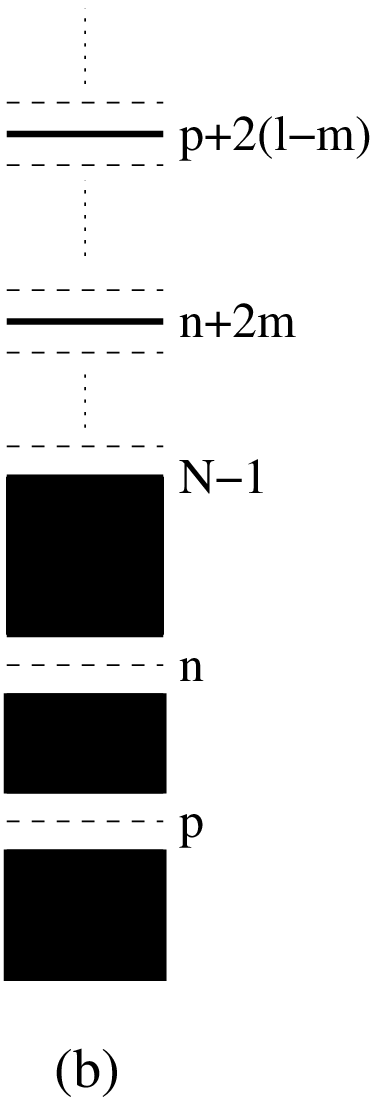} 
\hspace{3ex}
\includegraphics[height=7cm, width=3.5cm]{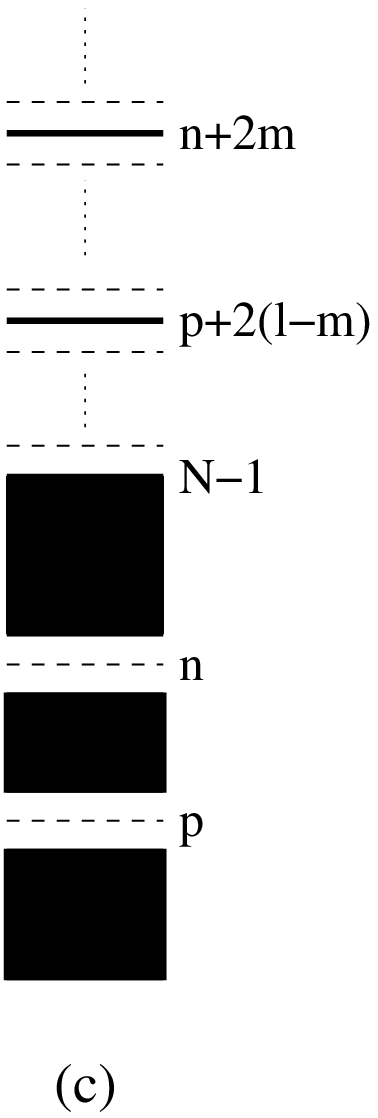} 
\caption{The states with two particle-hole pairs resulting from the action of two fermion bilinears on vacuum.}  
\label{fig:5} 
\end{figure} 
Adding up all the contributions, paying due attention to minus signs
coming from fermion anti-commutations, we get
\bea
\sum_{m=1}^{\frac{l-1}{2}}
\sum_{p=N+1-2(l-m)}^{N-1} && \biggl\{-\ket{4a}+\ket{4b}-\sum_{n=N+1-2m}^{p-2} \theta(2m+p-3-N)
\theta(p-2-n) \ket{5a} \nonumber \\
+\sum_{n=p+2}^{N-1} && \theta(2l+p-4m-2-n) \theta(n+2m-1-N) \theta(N-3-p) \ket{5b} \nonumber \\
-\sum_{n=p+2}^{N-1} && \theta(4m+n-2l-2-p) \theta(p+2l-2m-1-N) \theta(N-3-p) \ket{5c} 
\biggr\}. \nonumber \\  
\eea
In the above expression we have used the short-hand notation
$\ket{.}$ for the state shown in the figure with the corresponding
number. Also $\theta(n)=1$ for $n \geq 0$ and zero otherwise. Using
the rules of bosonization described in Section 2, one can write down
these states in the bosonic language. We get,
\bea
\sum_{m=1}^{\frac{l-1}{2}} \sum_{p=N+1-2l+2m}^{N-1} 
\biggl\{&-& (\sigmad_1)^{p+2l-2m-N} \sigmad_{N-p+2m} \ket{0} +
(\sigmad_1)^{p+2l-N} \sigmad_{N-p} \ket{0} \nonumber \\
&-&\sum_{n=N+1-2m}^{p-2} \theta(2m+p-3-N) \theta(p-2-n) \nonumber \\
&&~~~~~~~~\times (\sigmad_1)^{p+2l-4m-n-1} 
(\sigmad_2)^{n+2m-N} \sigmad_{N-p+1} \sigmad_{N-n} \ket{0} \nonumber \\
&+& \sum_{n=p+2}^{N-1} \theta(2l+p-4m-2-n) \theta(n+2m-1-N) \theta(N-3-p) \nonumber \\
&&~~~~~~~~\times (\sigmad_1)^{p+2l-4m-n-1} (\sigmad_2)^{n+2m-N} \sigmad_{N-n+1} 
\sigmad_{N-p} \ket{0} \nonumber \\
&-& \sum_{n=p+2}^{N-1} \theta(4m+n-2l-2-p) \theta(p+2l-2m-1-N) \theta(N-3-p) \nonumber \\
&&~~~~~~~~\times (\sigmad_1)^{n-2l+4m-p-1} (\sigmad_2)^{p+2l-2m-N} \sigmad_{N-n+1} 
\sigmad_{N-p} \ket{0} \biggr\}.
\eea
The contribution of the first two terms can be rewritten as 
\be
\sum_{m=1}^{\frac{l-1}{2}} \biggl(\sum_{k=1}^{l-m}-\sum_{k=m+1}^{l}\biggr)
(\sigmad_1)^{2l-2k+1} \sigmad_{2k-1} \ket{0},
\ee
which evaluates to precisely the right-hand side of
(\ref{identity}). To prove this identity, then, we need to show that
the contribution of the remaining three terms above vanishes. This
contribution can be rewritten as follows:
\bea
&& \sum_{m=1}^{\frac{l-1}{2}} \sum_{k} 
\biggl\{-\sum_{r=1}^{l-m}\sum_{s=r+1}^{m} 
\theta(m-r-1) \theta(s-r-1) \delta_{m, k+s} \nonumber \\
&+& \sum_{r=1}^{l-m}\sum_{s=1}^{r-1} \theta(l-m-s-k-1) 
\theta(s-2) \theta(m-r) \delta_{m, k+r} \nonumber \\
&-& \sum_{s=1}^{l-m}\sum_{r=1}^{s-1} \theta(l-m-r-k-1) \theta(s-2) 
\delta_{l-m, k+s} \biggr\}(\sigmad_1)^{2(l-r-s-2k)-1} (\sigmad_2)^{2k+1} 
\sigmad_{2r} \sigmad_{2s-1} \ket{0}. \nonumber \\
\eea
Notice that for a given value of $(k+s)$ either the first or the third
term contributes. This is because for odd $l$, $m$ and $(l-m)$ can
never be equal in the range $1 \leq m \leq \frac{l-1}{2}$. It is easy
to see that because of this the contribution of the middle term gets
cancelled for any given values of $k,~r,~s$. This proves the identity
(\ref{identity}). We have thus verified that the coefficients
$c^{l}_{\pm m}$ are given by (\ref{c}).

\section{Remarks on bosonization of nonrelativistic fermions in higher dimensions}

If we recall the definition of the exact fermi-bose equivalence
\cite{DMS}, it is clear that the fermionic oscillators $\psid_m,
\psi_n$ need not refer to fermions in one dimension, as long as they
satisfy the anticommutation relation 
\be
\{ \psid_m, \psi_n \}=\delta_{mn}, \; m,n \in Z^+ \equiv \{0,1,...,\infty\}
\label{1d}
\ee
E.g. consider fermions  moving
in a 2D harmonic oscillator potential. The fermionic oscillators
here can be labelled $\psid_{m_1, m_2}, \psi_{n_1, n_2}$ which
satisfy 
\be
\{\psid_{m_1, m_2}, \psi_{n_1, n_2} \} = \delta_{m_1, n_1}
  \delta_{m_2, n_2}, \;  m_1,m_2,n_1,n_2 \in Z^+
\label{2d}
\ee
Now it is easy to contruct (see below) invertible maps 
\be
f: Z^+ \times Z^+ \to Z^+, \;  m = f(m_1, m_2) 
\label{maps}
\ee 
Using such maps,
the ``2D'' fermion anticommutation relation \eq{2d} becomes
the ``1D'' relation \eq{1d}, with $m= f(m_1,m_2), n= f(n_1, n_2)$.

\def\floor#1{{\tt fl}\left(#1\right)}

The existence of invertible maps like
\eq{maps} follows from the countability of 
set $Z^+ \times Z^+$. An explicit construction of such a map is as follows. 
First, let us make a change of coordinates 
$(m_1,m_2) \mapsto (l, m_2)$ where
$l= m_1 + m_2$:  the range of $(l,m_2)$ are 
$l=0,1,2,...,\infty;~m_2=0,1,...,l $.
With this, the desired function $m=f(m_1,m_2)$ is defined as
\be
m= \mu(l,m_2) \equiv \frac{l(l+1)}2 + m_2
\ee
The inverse function $\mu \mapsto (m_1,m_2)$ is given by
\be
l(\mu) = \floor{\frac12(\sqrt{1+ 8\mu}-1)}; \, m_2(\mu)= \mu-
\frac12 l(\mu) \Big( l(\mu) +1 \Big)
\ee
The function $\floor{x}$ is defined as the largest integer contained
in the non-negative real number $x$.

Some examples of the values of $m_1,m_2$ and the corresponding $m$ are

\begin{tabular}{ l l l l l l l}
$(m_1, m_2)$ & (0,0) &(1,0) &(0,1) &(2,0) &(1,1) &(0,2)
\\
~~~~m & ~~~0  & ~~~1 & ~~~2 & ~~~3 & ~~~4 & ~~~5
\end{tabular}

Using this map, the fermionic hamiltonian for the 2-dimensional harmonic
oscillator, viz.
\bea
H_F &&= \sum_{m_1, m_2} {\cal E}_2(m_1,m_2) \psid(m_1,m_2) \psi(m_1,m_2), \; 
{\cal E}_2(m_1,m_2)\equiv m_1
+ m_2 + 1 = l+1
\nn
&&\equiv \sum_{\mu=0}^\infty {\cal E}_1(\mu)\psid_\mu \psi_\mu,
\;
{\cal E}_1(\mu) = l(\mu) + 1 \equiv \floor{\frac12(\sqrt{1+ 8\mu}-1)} + 1
\label{equiv-1d}
\eea
becomes 
\be
H_B = \sum_{k=1}^N {\cal E}_1(\hat n_k), \;
\hat n_k= \sum_{i=k}^N \ad_i a_i+N-k
\ee
Here ${\cal E}_1$ is the equivalent 1-dimensional fermionic energy level,
defined in \eq{equiv-1d}.

The above discussion proves that fermions in 2-dimensional harmonic oscillator
potential can be bosonized using our prescription. 
Similar remarks also apply to the case of fermions in a 2-dimensional box.

Indeed, fermions in an arbitrary D-dimensional harmonic oscillator
or a D-dimensional box can also be bosonized. For example, for a harmonic 
potential in D$=3$, the bosonic hamiltonian becomes
\bea
H_B &&= \sum_{k=1}^N  {\cal E}_1(\hat n_k),\;
\hat n_k = \sum_{i=k}^N \ad_i a_i+N-k
\nn
{\cal E}_1(\mu) &&= l(\mu) + \frac32,\; 
l_\mu = \floor{l^+(\mu)}
\eea 
where the function $l^+(\mu)$ is defined as the positive root of the cubic
equation $ l^+(\mu)(l^+(\mu)+1)(2 l^+(\mu) +4) - 12 \mu =0.$

Because of the appearance of the $\floor{...}$ function, the above
formulae, though exact, are not particularly easy to deal with. It
would be interesting to simplify these expressions by trying different
parameterizations of the $Z^+ \times ... Z^+$ lattice.

\underline{Dimension as an emergent concept}: In the above 
discussions of fermions in a D-dimensional harmonic oscillator, the 
dimension D can be read off from the equation for $E_1(\mu)$,
the equivalent 1D fermion. For $\mu \gg 1$, we have
\[
{\cal E}_1(\mu) \propto \mu^{1/D}
\]
This is easy to show, along the lines of the D=1,2,3. 
Note that this asymptotic formula is only valid for large $N$.
The dimensionality emerges only at large $N$.

     \newcommand{\sbibitem}[1]{\bibitem{#1}}
     
     \end{document}